\documentclass{amsart}

\usepackage{amssymb}

\title{Remarks on the group-theoretical foundations of 
particle physics}
\author{Robert Arnott Wilson}
\date{16th February 2020; revised 28th February 2020}
\newtheorem{remark}{Remark}
\newcommand{\Hom}{\mathrm{Hom}}
\address{School of Mathematical Sciences, Queen Mary University of London.}
\email{r.a.wilson@qmul.ac.uk}

\begin{document}
\begin{abstract}
I propose the group $SL(4,\mathbb R)$ as a generalisation of the
Dirac group $SL(2,\mathbb C)$ used in quantum mechanics,
as a possible basis on which to build a more general theory
from which the standard model of particle physics might be
derived as an approximation in an appropriate 
limit.
\end{abstract}

\maketitle
\section{Introduction}
The standard model of particle physics is based on four Lie groups,
that is the gauge groups $U(1)$, $SU(2)$ and $SU(3)$ of  electromagnetism,
and the weak and strong nuclear forces, respectively, together with the
Dirac group $SL(2,\mathbb C)$ that acts on the Dirac spinor and the
Dirac equation. All four of these groups are supposed to commute with each other,
but this is really only true in the high-energy limit. At practical energies
many of the symmetries implied by these groups are significantly `broken'.
This phenomenon is usually explained as having occurred immediately
after the Big Bang, as the energy density fell rapidly, in a process of
`spontaneous' symmetry-breaking.
In the case of the strong force, symmetry is restored by hypothesising abstract properties of
`colour', rather than seeking to model the three generations of fermions. The standard model
thus does not contain any symmetry group relating to the three generations.

Symmetry-breaking cannot be introduced into a commuting product
of groups by 
group-theoretical methods, and therefore 
geometrical methods are used in the standard
model. Group theory could only provide such a mechanism
if the groups do not commute with each other. It might then be possible to 
describe the effect of a change of energy scale on the parameters of the
standard model, provided only that the Dirac group fails to commute with
the relevant gauge groups.

In this paper I aim to provide a `proof of concept'
that a generalisation from commuting to non-commuting
groups has the potential to explain symmetry-breaking at a deeper
conceptual level than is possible in the standard model.
This includes a possible explanation for the three generations,
at least as far as electrons are concerned.
The aim is to do this as far as possible by changing the mathematical axioms,
with as little change to theorems or to physical applications as 
possible. 

\section{The ambient group}
The total real dimension of the three gauge groups is $12$. This extends to $18$
if we include the Dirac group as well. An alternative is to work over complex numbers, so that
the Dirac group is $3$-dimensional and the total dimension is $15$.
The eight smallest dimensions of complex simple Lie groups are listed in
Table~\ref{smalldim}. 

It is worth noting at this point that the group $SU(5)$
of type $A_4$ was used for the Georgi--Glashow Grand Unified Theory \cite{SU5GUT}
already in the 1970s. In restrospect, this group appears to have been too big,
in that it predicted new particles and new forces, that have not been detected experimentally.

The table therefore suggests that some group of type $A_3$ may be suitable, although there are
certainly other possibilities. The obvious group of type $A_3$ to try is $SL(4,\mathbb C)$,
although it is possible that some real form such as $SU(4)$ or $SL(4,\mathbb R)$ might be more appropriate.
Extensions to $U(4)$ or $GL(4,\mathbb R)$ or $GL(4,\mathbb C)$ are also potential
candidates. (Compare the Pati--Salam model \cite{PatiSalam}.)

\begin{table}
\caption{\label{smalldim}Small-dimensional complex Lie groups}
$$\begin{array}{rcccccccc}
\mbox{Type:}& A_1 & A_2 & B_2 & G_2 & A_3 & B_3/C_3 & A_4& D_4\cr
\mbox{Dimension:}& 3&8&10&14&15&21&24&28
\end{array}$$
\end{table}

For simplicity I shall work initially with $SL(4,\mathbb R)$. If any extensions to larger
groups seem to be required, these can be incorporated later on.
Note that $SL(4,\mathbb R)$ contains a subgroup of scalars of order $2$, and the 
quotient group is
\begin{eqnarray}
SL(4,\mathbb R)/Z_2 &\cong & SO(3,3)^\circ
\end{eqnarray}
that is, the connected component of the identity in the isometry group of a metric
\begin{eqnarray}
a^2+b^2+c^2-d^2-e^2-f^2
\end{eqnarray}
on a $6$-dimensional real space. 

This $6$-dimensional representation of $SO(3,3)^\circ$ is the fundamental bosonic
representation of $SL(4,\mathbb R)$, and can be constructed as the anti-symmetric square
of the defining representation on a $4$-dimensional space $V$. The latter, and its dual $V'$,
are the fundamental fermionic representations. Since this bosonic representation is self-dual,
it can also be described as the anti-symmetric square of the representation on $V'$.

\section{The Dirac gamma matrices}
The Dirac gamma matrices are a particular choice of basis for the algebra of $4\times 4$
complex matrices, specifically chosen to exhibit the structure of a complexified
Clifford algebra $\mathbb C\ell(1,3)$. They can also be used to generate the Lie algebra
$gl(4,\mathbb C)$, and specific choices among them generate the real Lie algebra
$sl(4,\mathbb R)$.

It is possible to produce a simple quaternionic notation for the Dirac matrices by
identifying $V$ with a copy of the quaternion algebra $\mathbb H$. First we construct
a Lie algebra $su(2)_R$ from right-multiplications by $i,j,k$. Then there is a
corresponding algebra $su(2)_L$ generated by left-multiplications by $-i,-j,-k$.
Let us write
\begin{eqnarray}
so(4) &=& su(2)_L\oplus su(2)_R\cr
&=& \langle i',j',k'\rangle \oplus \langle i,j,k\rangle
\end{eqnarray}
where we abuse notation to use $i,j,k$ for right-multiplications, and $i',j',k'$
for
left-multiplications by the quaternion conjugates $-i,-j,-k$.
Then $i,j,k$ commute with $i',j',k'$, so that the corresponding Lie brackets are $0$.
But in the ambient associative algebra we can also multiply these elements together in pairs.
Thus we obtain 
a total of $15$ linear maps that are easily seen to be linearly 
independent.

Since they have trace $0$, they span the Lie algebra $sl(4,\mathbb R)$.
They are listed in Table~\ref{sl4Rgens} together with the corresponding Dirac gamma matrices.
It is an easy exercise to check that the multiplication rules are the same in both notations.
For reference, a choice of matrix representation is also given in Table~\ref{matrices}.

\begin{table}
\caption{\label{sl4Rgens}Generators for the Lie algebra $sl(4,\mathbb R)$}
$$
\begin{array}{c|ccc|}
&i&j&k\cr\hline
i'&i'i&i'j&i'k\cr
j'&j'i&j'j&j'k\cr
k'&k'i&k'j&k'k\cr\hline
\end{array}\qquad
\begin{array}{c|ccc|}
& \gamma_2\gamma_3 & \gamma_3\gamma_1 & \gamma_1\gamma_2\cr\hline
i\gamma_5 & \gamma_0\gamma_1 & \gamma_0\gamma_2 & \gamma_0\gamma_3\cr
\gamma_5\gamma_0 & i\gamma_1 & i\gamma_2 & i\gamma_3\cr
i\gamma_0 & \gamma_1\gamma_5 & \gamma_2\gamma_5& \gamma_3\gamma_5\cr\hline
\end{array}
$$
\end{table}

\begin{table}
\caption{\label{matrices}Matrix representation of the Lie algebra}
$$\begin{array}{c|ccc|}
&\begin{pmatrix}.&+&.&.\cr -&.&.&.\cr .&.&.&-\cr .&.&+&.\end{pmatrix}
&\begin{pmatrix}.&.&+&.\cr .&.&.&+\cr -&.&.&.\cr .&-&.&.\end{pmatrix}
&\begin{pmatrix}.&.&.&+\cr .&.&-&.\cr .&+&.&.\cr -&.&.&.\end{pmatrix}\cr&&&\cr\hline
&&&\cr
\begin{pmatrix}.&-&.&.\cr +&.&.&.\cr .&.&.&-\cr .&.&+&.\end{pmatrix}
&\begin{pmatrix}+&.&.&.\cr .&+&.&.\cr .&.&-&.\cr .&.&.&-\end{pmatrix}
&\begin{pmatrix}.&.&.&-\cr .&.&+&.\cr .&+&.&.\cr -&.&.&.\end{pmatrix}
&\begin{pmatrix}.&.&+&.\cr .&.&.&+\cr +&.&.&.\cr .&+&.&.\end{pmatrix}\cr&&&\cr
\begin{pmatrix}.&.&-&.\cr .&.&.&+\cr +&.&.&.\cr .&-&.&.\end{pmatrix}
&\begin{pmatrix}.&.&.&+\cr .&.&+&.\cr .&+&.&.\cr +&.&.&.\end{pmatrix}
&\begin{pmatrix}+&.&.&.\cr .&-&.&.\cr .&.&+&.\cr .&.&.&-\end{pmatrix}
&\begin{pmatrix}.&-&.&.\cr -&.&.&.\cr .&.&.&+\cr .&.&+&.\end{pmatrix}\cr&&&\cr
\begin{pmatrix}.&.&.&-\cr .&.&-&.\cr .&+&.&.\cr +&.&.&.\end{pmatrix}
&\begin{pmatrix}.&.&-&.\cr .&.&.&+\cr -&.&.&.\cr .&+&.&.\end{pmatrix}
&\begin{pmatrix}.&+&.&.\cr +&.&.&.\cr .&.&.&-\cr .&.&-&.\end{pmatrix}
&\begin{pmatrix}+&.&.&.\cr .&-&.&.\cr .&.&-&.\cr .&.&.&+\end{pmatrix}\cr&&&\cr\hline
\end{array}$$
\end{table}

From this correspondence it is easy to see that the top two rows of the table generate a Lie subalgebra
$u(1)\oplus sl(2,\mathbb C)$. The bottom two rows form two copies of the representation of $sl(2,\mathbb C)$ 
on real $4$-dimensional Minkowski space.
Note however the `twisting' in the timelike coordinates, so that the two $4$-spaces are
\begin{eqnarray}
\langle j', k'i, k'j, k'k\rangle,\cr
\langle k',j'i,j'j,j'k\rangle.
\end{eqnarray}
Note also that the subalgebra $u(1)$ is generated not by a scalar $i$ (since the representation is real),
but by $i\gamma_5$. This fact has important consequences for the implementation
of the Dirac spinor in the proposed new notation.

\section{
Automorphisms and subgroups}
I have exhibited a copy of the Dirac group $SL(2,\mathbb C)$ inside $SL(4,\mathbb R)$, commuting
with the subgroup $U(1)$ of all elements of the form $\exp(i'x)$. In particular, $SL(2,\mathbb C)$ commutes with $i'$,
and therefore with the inner automorphism of $SL(4,\mathbb R)$ defined by conjugation by $i'$.

It turns out that the other groups we need, namely the gauge groups of the three forces,
can be obtained in a similar way from outer automorphisms.
The outer automorphism group of $SL(4,\mathbb R)$ is $Z_2\times Z_2$, generated by two
particular automorphisms that I shall call the \emph{chirality} and \emph{duality}
automorphisms.

The \emph{chirality} automorphism is defined by quaternion conjugation $q\mapsto \bar{q}$ on $V$. This map has
determinant $-1$, and swaps right-multiplication by $q$ with left-multiplication by $\bar{q}$. It therefore
transposes the multiplication table of the Clifford algebra, and centralizes the group $GL(3,\mathbb R)$,
and the associated Lie algebra $gl(3,\mathbb R)$ spanned by
\begin{eqnarray}
i'+i, j'+j, k'+k, & i'i, j'j, k'k, &i'j+j'i, j'k+k'j, k'i+i'k.
\end{eqnarray}
The $3\times 3$ scalar matrices in here are represented by 
elements of $\langle i'i+j'j+k'k\rangle$. Note also that the compact subalgebra is
\begin{eqnarray}
so(3) &=&\langle i'+i,j'+j,k'+k\rangle.
\end{eqnarray}

The \emph{duality} automorphism is defined by the transpose-inverse map on the group, or equivalently the
transpose-negative map on the Lie algebra. The fixed subalgebra is
\begin{eqnarray}
so(4) &=& su(2)_L\oplus su(2)_R\cr
&=& \langle i',j',k'\rangle \oplus \langle i,j,k\rangle.
\end{eqnarray}
This algebra also contains the subalgebra $so(3)$ just mentioned.

\begin{remark}
The product of the given chirality and duality automorphisms is a  chiral duality
automorphism with fixed subalgebra
\begin{eqnarray}
so(3,1)&=&\langle i'+i,j'+j,k'+k,i'j-j'i, j'k-k'j, k'i-i'k\rangle.
\end{eqnarray}
Notice in particular that
\begin{eqnarray}
so(3,1) &\cong& sl(2,\mathbb C),
\end{eqnarray}
but that the two copies of this Lie algebra exhibited here are not equal.
Indeed, they are disjoint subalgebras of $sl(4,\mathbb R)$. 

Moreover, the space $V$ supports a spinor representation of $sl(2,\mathbb C)$,
but a vector representation of $so(3,1)$.
This fact raises some interesting questions
of interpretation of these two subalgebras, and more particularly of the physical
interpretation of any isomorphism between them. This discussion is beyond the scope of this paper,
however, and will be presented elsewhere.
\end{remark}

To summarise, the chirality and duality automorphisms have fixed subalgebras $gl(3,\mathbb R)$ and
$so(4)$, which intersect in $so(3)$. If we restrict from $so(4)$ to $su(2)_L$, then we lose this intersection, and
obtain a direct sum of vector spaces
\begin{eqnarray}
gl(1,\mathbb R) + sl(3,\mathbb R) + su(2)_L.
\end{eqnarray}
This is not a direct sum of algebras, however, since $su(2)_L$ does not commute with
either $gl(1,\mathbb R)$ or $sl(3,\mathbb R)$. Correspondingly, the group $SU(2)_L$ does not
commute with $GL(1,\mathbb R)$ or $SL(3,\mathbb R)$.

If we were to work in the complex group $SL(4,\mathbb C)$ we could convert the split real forms
$GL(1,\mathbb R)$ and $SL(3,\mathbb R)$ into the compact real forms $U(1)$ and $SU(3)$.
By doing so we would obtain disjoint groups isomorphic to the gauge groups of the standard model.
It is, of course, not obvious that the groups so obtained bear any relationship to the actual gauge groups.
In order to try to throw more light on this question, I shall continue to work with the
split real forms, with this understanding that we can convert between different real forms
later on when/if necessary.

The important point to note at this stage is that the groups inside $SL(4,\mathbb R)$,
unlike the gauge groups of the standard model, do not commute with each other.
They therefore contain within themselves the seeds of symmetry-breaking, which might
therefore arise naturally from the group theory without having to be imposed from outside.

\section{Electroweak mixing}
The first test of this proposal is to see whether the mixing of electromagnetism with the weak force,
as described by the Glashow--Weinberg--Salam model \cite {GWS}, can be sensibly re-constructed within $SL(4,\mathbb R)$.
The four subalgebras mentioned above that may be involved in this process are
\begin{eqnarray}
gl(1,\mathbb R) &=& \langle i'i+j'j+k'k\rangle\cr
su(2)_L&=&\langle i',j',k'\rangle\cr
u(1)&=&\langle i'\rangle\cr
sl(2,\mathbb C) &=& \langle i,j,k,i'i,i'j,i'k\rangle.
\end{eqnarray}
The basic question to be resolved is how to mix the real scalar $i'i+j'j+k'k$ with the imaginary scalar $i'$.

In fact the above algebras $gl(1,\mathbb R)$ and $su(2)_L$ generate the full Lie algebra $sl(4,\mathbb R)$,
so it is worth first looking at the subalgebra generated by the two elements  $i'$ and $i'i+j'j+k'k$. A straightforward calculation
shows that this algebra is $5$-dimensional, and breaks up as a direct sum of three subalgebras:
\begin{eqnarray}
gl(1,\mathbb R) &=& \langle i'i\rangle\cr
u(1)&=& \langle i'+i\rangle\cr
sl(2,\mathbb R)_W&=&\langle i'-i, j'j+k'k, j'k-k'j\rangle.
\end{eqnarray}
It is possible to combine the algebra $gl(1,\mathbb R)$ with either of the other two, and express the $5$-dimensional
algebra in one of the two forms
\begin{eqnarray}
gl(1,\mathbb C) \oplus sl(2,\mathbb R)&\cong&
u(1) \oplus gl(2,\mathbb R).
\end{eqnarray}

This subalgebra suggests that the mixing may come in two parts, one of
which is 
a mixing of the chiral $u(1)$ generated by $i'-i$ with
the non-chiral $u(1)$ generated by $i'+i$.
The other is a mixing of the two real scalars $i'i$ and $j'j+k'k$.

In particular, the algebra $su(2)_L$ generated by $i',j',k'$ is modified by
\begin{itemize}
\item mixing $i'$ with the element $i$ from $su(2)_R$, and
\item mixing $j'$ and $k'$ as $j'+k'i$ and $j'i+k'$, with a further factor of $j$ on the right and left respectively.
\end{itemize}
This mathematical formalism is almost identical to that
used in the standard model to describe electroweak mixing. But instead of being imposed from the outside,
it arises naturally from the embeddings of the various algebras in $sl(4,\mathbb R)$. 

More specifically, the action of 
$sl(2,\mathbb R)_W$ on $\mathbb H$ is given by
\begin{eqnarray}
i'-i &:& (1,i,j,k) \mapsto(-i,1,0,0)\cr
j'j+k'k&:& (1,i,j,k)\mapsto (1,-i,0,0)\cr
j'k-k'j&:& (1,i,j,k)\mapsto (-i,-1,0,0)
\end{eqnarray}
so that it acts only on the `left-handed' part of the spinor in the $1,i$ coordinates, not on the
`right-handed' part in the $j,k$ coordinates.
There is a corresponding `right-handed' copy of $sl(2,\mathbb R)$ given by changing the signs, that lies inside $sl(3,\mathbb R)$,
and is therefore (conjecturally) related to the strong force:
\begin{eqnarray}
sl(2,\mathbb R)_S &=& \langle i'+i, j'j-k'k, j'k+k'j\rangle
\end{eqnarray}
The two copies of $sl(2,\mathbb R)$ commute with each other, and with $i'i$. 
The corresponding group is 
\begin{eqnarray}
\mathbb R^\times_{>0} \times SL(2,\mathbb R)_W \times SL(2,\mathbb R)_S &\cong& \mathbb R^\times_{>0} \times Spin(2,2).
\end{eqnarray}

\begin{remark}
This group $Spin(2,2)$ is 
not to be confused with the subgroup $SO(2,2)$, 
although their Lie algebras are isomorphic. 
For example, we may take a metric in which $1$ and $i$ have norm $1$ and
$j$ and $k$ have norm $-1$, so that
\begin{eqnarray}
so(2,2) &=& \langle i, i', i'j, i'k, j'i, k'i\rangle\cr
&=& \langle i,i'j,i'k\rangle \oplus \langle i',j'i, k'i\rangle\cr
&=& sl(2,\mathbb R)_A \oplus sl(2,\mathbb R)_B.
\end{eqnarray}
In this case the corresponding groups $SL(2,\mathbb R)$ both contain the scalar matrix $-1$, so their product is
not a direct product but a central product
\begin{eqnarray}
SO(2,2) &\cong& SL(2,\mathbb R)_A\circ SL(2,\mathbb R)_B.
\end{eqnarray}
This group 
has no obvious application in particle physics, as far as I know. 
\end{remark}

\begin{remark}
The five groups $SO(4)$, $SO(3,1)$, $SO(2,2)$, $Spin(3,1)$  and $Spin(2,2)$ are, up to conjugacy, 
the only subgroups of $SL(4,\mathbb R)$
that are real forms of the Lie group of type $A_1A_1$. 
Of these groups, only $Spin(2,2)$ splits $V$ into two halves that could be identified with the left- and right-handed
spinors in the standard model.
\end{remark}

\begin{remark}
If we identify the complex structure defined by $i$ in the Clifford algebra notation, with that defined by $i$ in the
quaternionic notation, then the effect on the Clifford algebra is to identify $i$ with $\gamma_2\gamma_3$.
Of course, $\gamma_2\gamma_3$ does not commute with all of the Clifford algebra, so one has to take care
that the multiplication is always done on the same side. We obtain an identification of $\gamma_5$ with
$\gamma_0\gamma_1$, which has eigenvalues $1,1,i,-i$, so that $1-\gamma_5$ becomes a projection onto
half the space as in the standard model.

The subalgebra $sl(2,\mathbb R)_W$ is likewise spanned by
\begin{eqnarray}
\gamma_2\gamma_3(\gamma_5-1), & \gamma_2(\gamma_5-1), & \gamma_3(\gamma_5-1),
\end{eqnarray}
so that it behaves as an algebra that acts only on half of the spinor. In the proposed notation this half is a
real $2$-space, but since the required eigenvalues are complex, one needs to extend it to a complex
$2$-space in order to match the standard model.
\end{remark}

\section{The strong force}
The above remark involves a choice of a particular direction in the Clifford algebra (defined by $i'i$) in which to define
or measure spin. This choice splits the space $V$ into two $2$-dimensional eigenspaces.
By symmetry it is necessary to replicate the construction for $j'j$ and $k'k$. This gives three copies of a $5$-dimensional
algebra, which would be enough to cover the whole of $sl(4,\mathbb R)$, were it not for the fact that
the terms $i'i,j'j,k'k$ are double-counted. We are therefore missing three dimensions, which are spanned by
the following elements,
that are chosen to be perpendicular to all three copies of
$gl(1,\mathbb C)\oplus sl(2,\mathbb R)$, with respect to the Killing form:
\begin{eqnarray}
i'j+j'i, j'k+k'j, k'i+i'k.
\end{eqnarray}

Now a straightforward calculation shows that
the Lie algebra generated by these three extra elements is $sl(3,\mathbb R)$,
spanned by elements like the following (with irrelevant scalar factors suppressed):
\begin{eqnarray}
[j'k+k'j,k'i+i'k]&=&k'+k,\cr
[k'+k,i'j+j'i]&=&j'j-i'i.
\end{eqnarray}
These elements appear in the three copies of $gl(1,\mathbb R)\oplus u(1)\cong gl(1,\mathbb C)$
already discussed in connection with the electroweak forces:
\begin{eqnarray}
\label{C3}
&\langle i'i, i'+i\rangle,\cr
&\langle j'j, j'+j\rangle,\cr
&\langle k'k, k'+k\rangle.
\end{eqnarray}

If all these conjectural interpretations are valid, then these three complex numbers
can be used to describe some mixing of the
strong force with the electroweak forces.
More specifically, there is a map from one copy of $\mathbb C^3$ with a basis suited to the strong force, to
another copy, with a basis suited to the weak force. Such a map can be written as a $3\times 3$ complex
matrix. 
It is therefore possible to insert the Cabibbo--Kobayashi--Maskawa (CKM) matrix 
\cite{Cabibbo,KM} in this place in
the model.

In this way the model is able (at least in principle)
to describe the mixing of the strong force with the electroweak forces.
Since the CKM matrix describes the mixing of quark generations relative to lepton generations,
the generations are described in the proposed model by three copies of the complex numbers, that
in the lepton case may
be taken  to be as in (\ref{C3}).

\section{The Georgi--Glashow model}
Another subgroup of $SL(4,\mathbb R)$ that may be of interest is that obtained from $SL(2,\mathbb C)$ by
adjoining one of the two vector representations. This subgroup is isomorphic to $Sp(4,\mathbb R)$.
The corresponding Lie algebra is generated by the Dirac matrices $i\gamma_\mu$, for 
$i=0,1,2,3$, or in quaternionic notation, by $k',j'i,j'j,j'k$. The ten dimensions of the
algebra are given by
\begin{eqnarray}
i,j,k,k',i'i,i'j,i'k,j'i,j'j,j'k
\end{eqnarray}
and the remaining five dimensions form the vector representation of $so(2,3)\cong sp(4,\mathbb R)$:
\begin{eqnarray}
i',j',k'i,k'j,k'k.
\end{eqnarray}

The structure revealed by this group is reminiscent of that used in the Georgi--Glashow model
based on $SU(5)$. One difference is that the new model is real rather than complex, so that the
$24$-dimensional group $SU(5)$ is replaced by the $10$-dimensional group $SO(2,3)$. This reduction
in dimension could perhaps allow for a version of the Georgi--Glashow model that does not
predict new forces or new particles. Another difference is that the $2+3$ splitting appears
naturally rather than having to be imposed.

The $15$ fundamental fermions of a single generation could then be allocated to the Lie algebra
by using $i,j,k$ for the three colours, and $i',j',k'$ for the leptons. One might for example label the algebra
thus:
\begin{eqnarray}
\begin{array}{c|ccc|}
& u^r_R & u^g_R & u^b_R\cr\hline
\nu_e & u^r_L & u^g_L & u^b_L\cr
e_L & d^r_L & d^g_L & d^b_L\cr
e_R & d^r_R& d^g_R & d^b_R\cr\hline
\end{array}
\end{eqnarray}
The other two generations could be labelled in a similar manner, using copies of $Sp(4,\mathbb R)$
defined by singling out $i'$ or $j'$ rather than $k'$ from $su(2)_L$. 
Using the Lie algebra rather than the Clifford algebra to describe these particles has the possible
advantage that right-handed neutrinos, which have not been observed experimentally, do not appear.

%
On the other hand, the fact that the Georgi--Glashow model does not address the question of the
three generations implies that there may not be a close connection with the proposed
use of $SL(4,\mathbb R)$.
Further exploration of the fermionic representations of $SL(4,\mathbb R)$ 
will be undertaken in a forthcoming paper.

\section{A heuristic mass equation}
The proposed model allows (or requires) basic properties of fermions to
be modelled in small fermionic representations of $SL(4,\mathbb R)$. The first generation of
leptons is related to the operations $i'i$ and $i'+i$ that describe commutation by $i$ in the
group-theoretical sense and the Lie algebra sense respectively:
\begin{eqnarray}
i'i &:& g \mapsto -igi=i^{-1}gi\cr
i'+i&:& a \mapsto -ia+ai = [a,i].
\end{eqnarray}
The second and third generations of leptons are described by the corresponding maps for $j$ and $k$.
Abstracting these properties to a $4$-dimensional representation on $\mathbb H$ allows us to
represent the $3$ generations by vectors $i$, $j$ and $k$ respectively. The real part $1$ then appears to
describe the charge of the leptons. 
On the other hand, $i,j,k$ as right-multiplications appear in the model as colours of quarks,
from which one can coordinatise protons and neutrons with all three coordinates $i,j,k$.

Hence we obtain the following (conjectural) descriptions of the three generations of leptons,
together with the two lightest baryons:
$$\begin{array}{clccl}
e: &(-1,1,0,0) &\quad& \nu_e:& (0,1,0,0)\cr
\mu: & (-1,0,1,0) &&\nu_\mu: & (0,0,1,0)\cr
\tau: &(-1,0,0,1) && \nu_\tau: & (0,0,0,1)\cr
p: & (1,1,1,1) && n:& (0,1,1,1)
\end{array}$$
 
 These coordinates are intended to describe intrinsic properties of the particles,
 in a copy of $V$. The particles are then
 embedded in spacetime, described by another copy of $V$. Measurement of
 (externally visible) properties of the particles might then be obtained via a mapping from internal spacetime to
 external spacetime. These mappings can be expressed in the representation
 \begin{eqnarray}
 \Hom(V,V) &\cong& V' \otimes V,
 \end{eqnarray}
 which breaks up as a scalar plus the adjoint representation.
 I suggest that the scalar measures the mass, while the adjoint representation contains
 information about various other properties such as spin.

 If one adds together the $6$ leptons and three protons,
 taking into account that the proton contains three colours of quarks, one sees a total of $15$ dimensions
 of fundamental properties, encoded in the vector
 \begin{eqnarray}
 &&(0,5,5,5).
 \end{eqnarray}
These $15$ dimensions can be re-distributed by changing basis on the adjoint representation,
and can apparently be allocated instead to five neutrons. Such a re-distribution preserves the overall
charge, which I have (conjecturally) put in the real part of $\mathbb H$, and therefore can be achieved by an element of $sl(3,\mathbb R)$.
This group corresponds to an action of the strong force, which in the standard model does not
change particle masses. Therefore the model suggests that the total mass of the six leptons and
three protons should be equal to the total mass of five neutrons. 

Since the neutrinos have negligible mass compared to all the other particles in this equation,
the effective prediction is that 
\begin{eqnarray}
\label{emutau}
m(e)+m(\mu)+m(\tau)+3m(p) &=& 5m(n).
\end{eqnarray}
Of these masses, the $\tau$ mass is the
least accurately known, by about three orders of magnitude. Hence
we can re-cast this equation as a prediction
of the $\tau$ mass.

I take
experimental values in $\mathrm{MeV}/c^2$ from CODATA 2014 \cite{CODATA} as
\begin{eqnarray}
m(e) &=& .5109989461(31)\cr
m(\mu) &=&105.6583745(24)\cr
m(p) &=& 938.2720813(58)\cr
m(n) &=& 939.5654133(58).
\end{eqnarray}
Then the
prediction is
\begin{eqnarray}
m(\tau) _p 
&=& 1776.84145(3).
\end{eqnarray}
This gives a prediction to $8$ significant figures, 
 well within current 
experimental 
uncertainty, given by 
\begin{eqnarray}
m(\tau)_e=1776.86(12). 
\end{eqnarray}

\section{Another mass equation}
One can apply a similar analysis to the fundamental bosonic representation of dimension $6$.
In this case the six coordinates appear in two sets of three, each set associated in some way to
the triple $i,j,k$. Again it looks as though there is a distinction between charge and spin,
though in this case we appear to require three charges (say $+$, $-$ and $0$) and three
spins (no longer directly associated to specific generations of leptons). 
The three fundamental massive bosons are the intermediate vector bosons, that is the
bosons $Z^0$, $W^+$ and $W^-$ that mediate the weak force. 

In this case we have that $\Lambda^2(V)$ is self-dual, and therefore
\begin{eqnarray}
\Hom(\Lambda^2(V),\Lambda^2(V)) &\cong & \Lambda^2(V) \otimes \Lambda^2(V)\cr
&\cong& \Lambda^2(\Lambda^2(V))\oplus S^2(\Lambda^2(V)).
\end{eqnarray}
Now $\Lambda^2(\Lambda^2(V))$ is again the adjoint representation, while
$S^2(\Lambda^2(V))$ breaks up as a scalar plus a $20$-dimensional irreducible representation
that is used in the different context of general relativity for the Riemann Curvature Tensor.

The intermediate vector bosons appear to be described by $su(2)_L$ together with three copies of $sl(2,\mathbb R)_W$,
and therefore to need a total of $12$ fundamental spins and charges. This can be achieved by allocating each of them
two fermionic spins and two fermionic charges. In this way they have spin $1$, and total charge either $1+0$, $-1+0$ or $-1+1$.
In the $6$-dimensional representation, therefore, we can choose coordinates such that the particles are (conjecturally)
represented as follows:
\begin{eqnarray}
Z^0&=& (0,1,1;0,1,1)\cr
W^+&=&(1,1,0;1,1,0)\cr
W^-&=&(1,0,1;1,0,1).
\end{eqnarray}

Adding up the three vectors of quantum numbers, we obtain $(2,2,2;2,2,2)$, which can separated into a
spin $0$ component $(2,2,2;0,0,0)$ and the remainder $(0,0,0;2,2,2)$. The spin $0$ component must surely be
two copies of the Higgs boson
\begin{eqnarray}
H^0&=&(1,1,1;0,0,0).
\end{eqnarray}
The remainder consists of two copies of the fermionic particle $(0,0,0;1,1,1)$, which was earlier identified as a neutron.
Hence the model suggests the following equation: 
\begin{eqnarray}
\label{Higgseqn}
2m(H^0) + 2m(n) &=& m(Z^0) + m(W^+) +m(W^-).
\end{eqnarray}

From a historical perspective, at least, it makes sense to
regard this equation as a prediction of the mass of the Higgs boson in terms of previously
discovered particles.  If we 
substitute 
experimental values, in $\mathrm{MeV}/c^2$, of
\begin{eqnarray}
m(Z^0) &=& 91187.6(2.1)\cr
m(W^\pm)&=& 80379(12)\cr
m(n)&\approx& 939.6
\end{eqnarray}
then we
obtain the prediction (or post-diction)
\begin{eqnarray}
m(H^0)_p&=& 125033(12),
\end{eqnarray}
compared 
to the current 
experimental value of 
\begin{eqnarray}
m(H^0)_e &=& 125180(160).
\end{eqnarray}
The prediction is accurate to $1\sigma$,
and predicts one more significant figure. 
Incidentally, without the small contribution of the neutron, required by the model,
the calculation of the Higgs mass 
would differ from experiment by some 
$5\sigma$.

\section{Conclusion}
In this paper I have shown how all the groups that appear in the foundation of the standard model of particle physics
arises naturally from the group $SL(4,\mathbb R)$, and the associated Lie algebra $sl(4,\mathbb R)$.
In particular, the coupling of mass to charge in a model based on this algebra
seems to require the existence of three generations of leptons, and
the existence of a chiral force that acts on the three generations, and has a gauge group
that is either $SU(2)$ or $SL(2,\mathbb R)$. 
Furthermore, this chiral
force does not exhaust the Lie algebra, which contains three more dimensions that generate a Lie algebra
$sl(3,\mathbb R)$, representing a non-chiral force that can plausibly be identified with the strong force.

Some mixing of the forces arises naturally from the group theory, and is an inevitable consequence of the fact that the
groups that take the place of the gauge groups do not commute with each other.
This mixing therefore occurs without the need to hypothesise any
`spontaneous' symmetry-breaking in the Big Bang. While I have not explained the values of the mixing
parameters in this paper, I have explained the number of them, and their mathematical structure.

In other words, many of the puzzling but essential ingredients of the standard model arise as an inevitable consequence of
the algebraic structure of $4$-dimensional space, 
and in particular of a group $SL(4,\mathbb R)$ of local
symmetries of space, 
interpreted in the standard model as internal symmetries of elementary particles themselves.

Of course, I have not constructed a complete model here. More work will be required to verify that the
details of the standard model can be consistently included, and not just the broad structures.
For example, it is not even clear that the proposals put forward in this paper are consistent
with the basic axioms of quantum field theory. It remains to be seen whether this
problem can be overcome.

As an illustration of the power of the proposed new foundation for the standard model,
I have provided a heuristic justification for two conjectured 
mass equations which are predictive, and which can be tested by measuring the
masses 
of the tau lepton and the Higgs boson 
to greater accuracy than is currently known.
One of these gives a possible hint as to how the differing masses of the three generations might arise.

\section*{Acknowledgements}
The author would like to thank the Isaac Newton Institute for Mathematical Sciences, Cambridge,
for support and hospitality during the programme `Groups, representations and applications: new perspectives'
where work on this paper was undertaken. This work was supported by EPSRC grant no.
EP/R014604/1.

\end{document}